\documentclass[12pt]{article}

\usepackage{boxedminipage}
\usepackage[english]{babel}
\usepackage{latexsym}
\usepackage{amssymb}
\usepackage{times}

\usepackage{amsmath}

\newtheorem{theorem}{Theorem}[section]

\newtheorem{definition}[theorem]{Definition}

\newcounter{itemcount}

\begin{document}

\title{Resettable Zero Knowledge in the Bare Public-Key Model under Standard Assumption}

\author{Yi Deng, \ Dongdai Lin\\
\small   The state key laboratory of information
security, Institute of software,\\
\small   Chinese Academy of sciences, Beijing, 100080, China\\
\small   Email: {\tt \{ydeng,ddlin\}@is.iscas.ac.cn } }
\date{}
\maketitle

\begin{abstract}
In this paper we resolve an open problem regarding resettable zero
knowledge in the bare public-key (BPK for short) model: Does there
exist constant round resettable zero knowledge argument with
concurrent soundness for $\mathcal{NP}$ in BPK model without
assuming \emph{sub-exponential hardness}? We give a positive answer
to this question by presenting such a protocol for any language in
$\mathcal{NP}$ in the bare public-key model assuming only
collision-resistant hash functions against \emph{polynomial-time}
adversaries. 
\end{abstract}

\section{Introduction}
Zero knowledge (ZK for short) proof, a proof that reveals nothing
but the validity of the assertion, is put forward in the seminal
paper of Goldwasser, Micali and Rackoff \cite{GMR}. Since its
introduction, especially after the generality demonstrated in
\cite{GMW}, ZK proofs have become a fundamental tools in design of
some cryptographic protocols. In recent years, the research is
moving towards extending the security to cope with some more
malicious communication environment. In particular, Dwork et al.
\cite{DNS}introduced the concept of concurrent zero knowledge, and
initiate the study of the effect of executing ZK proofs concurrently
in some realistic and asynchronous networks like the Internet.
Though the concurrent zero knowledge protocols have wide
applications, unfortunately, they requires logarithmic rounds for
languages outside $\mathcal{BPP}$ in the plain model for the
black-box case \cite{CKPR} and therefore are of round inefficiency.
In the Common Reference String model, Damgaard \cite{D1} showed that
3-round concurrent zero-knowledge can be achieved efficiently.
Surprisingly, using non-black-box technique, Barak \cite{B}
constructed a constant round non-black-box bounded concurrent zero
knowledge protocol though it is very inefficient.

Motivated by the application in which the prover (such as the user
of a smart card) may encounter resetting attack, Canetti et al.
\cite{CGGM} introduced the notion of resettable zero knowledge (rZK
for short). An rZK formalizes security in a scenario in which the
verifier is allowed to reset the prover in the middle of proof to
any previous stage. Obviously the notion of resettable zero
knowledge is stronger than that of concurrent zero knowledge and
therefore we can not construct a constant round black-box rZK
protocol in the plain model for non-trivial languages. To get
constant round rZK, the work \cite{CGGM} also introduced a very
attracting model, the bare public-key model(BPK). In this model,
Each verifier deposits a public key $pk$ in a public file and stores
the associated secret key $sk$ before any interaction with the
prover begins. Note that no protocol needs to be run to publish
$sk$, and no authority needs to check any property of $pk$.
Consequently the BPK model is considered as a very weak set-up
assumption compared to previously models such as common reference
model and PKI model.

However, as Micali and Reyzin \cite{MR1} pointed out, the notion of
soundness in this model is more subtle. There are four distinct
notions of soundness: one time, sequential, concurrent and
resettable soundness, each of which implies the previous one.
Moreover they also pointed out that there is NO black-box rZK
satisfying resettable soundness for non-trivial language and the
original rZK arguments in the BPK model of \cite{CGGM} does not seem
to be concurrently sound. The 4-round(optimal) rZK arguments with
concurrent soundness in the bare public-key model was proposed by Di
Crescenzo et al. in \cite{DPV2} and also appeared in \cite{YZ}.

 All above rZK arguments in BPK model
  need some cryptographic primitives secure
 against sub-exponential time adversaries, which is not a standard
 assumption in cryptography. Using non-black-box techniques, Barak
 et al. obtained a constant-round rZK argument of knowledge assuming only
 collision-free hash functions secure against
 supperpolynomial-time algorithms\footnote{using idea from\cite{BG},
 this results also holds under standard assumptions that there
 exist hash functions that are collision-resistent against all
 polynomial-time adversaries.}, but their protocol enjoys only
 sequential soundness. The existence of constant round rZK arguments
 with concurrent soundness in BPK model under only polynomial-time
 hardness assumption is an interesting problem.

 \bigskip\noindent\textbf{Our results.}
 In this paper we resolve the above open problem by presenting a constant-round rZK argument
 with concurrent soundness in BPK model for $\mathcal{NP}$ under the
 standard assumptions that there exist hash functions collision-resistant
 against \emph{polynomial time} adversaries, We note that our
 protocol is a argument of knowledge and therefore the
 non-black-box technique is inherently used.

  In our protocol, we use the resettably-sound non-black-box zero knowledge argument
as a building block in a manner different from that in \cite{BGGL}:
instead of using it for the verifier to prove the knowledge of its
secret key, the verifier uses it in order to proves that a challenge
matches the one he committed to in a previous step. This difference
is crucial in the concurrent soundness analysis of our protocol: we
just need to simulate \emph{only one execution} among all concurrent
executions of the resettably-sound zero knowledge argument for
justifying concurrent soundness, instead of simulating all these
concurrent executions.
% (so we avoids the big problem: how to
%construct constant round concurrent ZK arguments).

\section{Preliminaries}
In this section we recall some definitions and tools that will be
used later.

 In the following we say that function $f(n)$ is
negligible if for every polynomial $q(n)$ there exists an $N$ such
that for all $n\geq N$, $f(n)\leq 1/q(n)$. We denote by
$\delta\leftarrow_{\small{R}}\Delta$ the process of picking a random
element $\delta$ from $\Delta$.

\bigskip\noindent\textbf{The BPK Model}.The bare public-key model(BPK
model)assumes that:
\begin{itemize}
    \item A public file $F$ that is a collection of records, each
containing a verifier's public key, is available to the prover.
    \item An (honest)prover $P$is an interactive deterministic
    polynomial-time algorithm that is given as inputs a secret
    parameter $1^n$, a $n$-bit string $x\in L$, an auxiliary input
    $y$, a public file $F$ and a random tape $r$.
    \item An (honest) verifier $V$ is an interactive deterministic
    polynomial-time algorithm that works in two stages. In stage
    one, on input a security parameter $1^n$ and a random tape
    $w$, $V$ generates a key pair $(pk, sk)$ and stores $pk$ in
    the file $F$. In stage two, on input $sk$, an $n$-bit string
    $x$ and an random string $w$, $V$ performs the interactive
    protocol with a prover, and outputs "accept $x$" or "reject
    $x$".
\end{itemize}

\begin{definition} We say that the protocol $<P,V>$ is
complete for a language $L$ in $\mathcal{NP}$, if for all $n$-bit
string $x\in L$ and any witness $y$ such that $(x,y)\in R_L$, here
$R_L$ is the relation induced by $L$, the probability that $V$
interacting with $P$ on input $y$, outputs "reject $x$" is
negligible in $n$.
\end{definition}

\textbf{Malicious provers and Its attacks in the BPK model}. Let
$s$ be a positive polynomial and $P^*$ be a probabilistic
polynomial-time algorithm on input $1^n$.

$P^*$ is a \emph{s-concurrent malicious} prover if on input a
public key $pk$ of $V$, performs at most s interactive protocols
as following: 1) if $P^*$ is already running $i-1$ interactive
protocols $1\leq i-1\leq s$, it can output a special message
"Starting $x_i$," to start a new protocol with $V$ on the new
statement $x_i$; 2) At any point it can output a message for any
of its interactive protocols, then immediately receives the
verifier's response and continues.

A concurrent attack of a \emph{s-concurrent malicious} prover
$P^*$ is executed in this way: 1) $V$ runs on input $1^n$ and a
random string and then obtains the key pair $(pk,sk)$; 2) $P^*$
runs on input $1^n$ and $pk$. Whenever $P^*$ starts a new protocol
choosing a statement, $V$ is run on inputs the new statement, a
new random string and $sk$.

\begin{definition} $<P,V>$ satisfies \emph{concurrent
soundness} for a language $L$ if for all positive polynomials $s$,
for all \emph{s-concurrent malicious} prover $P^*$, the
probability that in an execution of concurrent attack, $V$ ever
outputs "accept $x$" for $x\notin L$ is negligible in $n$.
\end{definition}

The notion of resettable zero-knowledge was first introduced in
\cite{CGGM}. The notion gives a verifier the ability to rewind the
prover to a previous state (after rewinding the prover uses the
same random bits), and the \emph{malicious} verifier can generate
an arbitrary file $F$ with several entries, each of them contains
a public key generated by the malicious verifier. We refer readers
to that paper for intuition of the notion. Here we just give the
definition.

\begin{definition} An interactive argument system $<P,V>$ in
the BPK model is black-box resettable zero-knowledge if there
exists a probabilistic polynomial-time algorithm $S$ such that for
any probabilistic polynomial-time algorithm $V^*$, for any
polynomials $s$, $t$, for any $x_i\in L$, the length of $x_i$ is
$n$, $i=1,...,s(n)$, $V^*$ runs in at most $t$ steps and the
following two distributions are indistinguishable:
\begin{enumerate}
    \item the view of $V^*$ that generates $F$ with $s(n)$
    entries and interacts (even concurrently) a polynomial number
    of times with each $P(x_i,y_i,j,r_k,F)$ where $y_i$ is a
    witness for $x_i\in L$, $r_k$ is a random tape and $j$ is the
    identity of the session being executed at present for $1\leq i,j,k \leq
    s(n)$;
    \item the output of $S$ interacting with on input
    $x_1,...x_{s(n)}$.
\end{enumerate}
\end{definition}

 \bigskip\noindent\textbf{$\Sigma\textbf{-protocols}$} A protocol $<P,V>$ is said
to be $\Sigma$-protocol for a relation $R$ if it is of 3-move form
and satisfies following conditions:\begin{enumerate}
    \item \emph{Completeness}: for all $(x,y)\in R$, if $P$ has
    the witness $y$ and follows the protocol, the verifier always
    accepts.
    \item \emph{Special soundness}: Let $(a,e,z)$ be the three
    messages exchanged by prover $P$ and verifier $V$. From any
    statement $x$ and any pair of accepting transcripts $(a,e,z)$
    and $(a,e',z')$ where $e\neq e'$, one can efficiently compute
    $y$ such that $(x,y)\in R$.
    \item \emph{Special honest-verifier ZK}: There exists a
    polynomial simulator $M$, which on input $x$ and a random $e$
    outputs an accepting transcript of form $(a,e,z)$ with the
    same probability distribution as a transcript between the
    honest $P$, $V$ on input $x$.
\end{enumerate}
Many known efficient protocols, such as those in \cite{GQ} and
\cite{S}, are $\Sigma$-protocols. Furthermore, there is a
$\Sigma$-protocol for the language of Hamiltonian Graphs \cite{B},
assuming that one-way permutation families exists; if the commitment
scheme used by the protocol in \cite{B} is implemented using the
scheme in \cite{N} from any pseudo-random generator family, then the
assumption can be reduced to the existence of one-way function
families, at the cost of adding one preliminary message from the
verifier. Note that adding one message does not have any influence
on the property of $\Sigma$-protocols: assuming the new protocol is
of form $(f,a,e,z)$, given the challenge $e$, it is easy to
indistinguishably generate the real transcript of form $(f,a,e,z)$;
given two accepting transcripts $(f,a,e,z)$ and $(f,a,e',z')$, where
$e\neq e'$, we can extract a witness easily. We can claim that any
language in $\mathcal{NP}$ admits a 4-round $\Sigma$-protocol under
the existence of any one-way function family (or under an
appropriate number-theoretic assumption), or a $\Sigma$-protocol
under the existence of any one-way permutation family. Though the
following OR-proof refers only to 3-round $\Sigma$-protocol, readers
should keep in mind that the way to construct the OR-proof is also
applied to 4-round $\Sigma$-protocol.

Interestingly, $\Sigma$-protocols can be composed to proving the OR
of atomic statements, as shown in \cite{DDPY,CDS}. Specifically,
given two protocols $\Sigma_0$,$\Sigma_1$ for two relationships
$R_0$, $R_1$, respectively, we can construct a
$\Sigma_{OR}$-protocol for the following relationship efficiently:
$R_{OR}={((x_0,x_1),y): (x_0,y)\in {R_0} or (x_1,y)\in {R_1}}$, as
follows. Let $(x_b,y)\in {R_b}$ and $y$ is the private input of $P$.
$P$ computes $a_b$ according the protocol $\Sigma_b$ using $(x_b,
y)$. $P$ chooses $e_{1-b}$ and feeds the simulator $M$ guaranteed by
$\Sigma_{1-b}$ with $e_{1-b}, x_{1-b}$, runs it and gets the output
$(a_{1-b},e_{1-b},z_{1-b})$. $P$ sends $a_b$, $a_{1-b}$ to $V$ in
first step. In second step, $V$ picks $e\leftarrow_{\small{R}}
\mathbb{Z}_q$ and sends it to $P$. Last, $P$ sets $e_b=e\oplus
e_{1-b}$, and computes the last message $z_b$ to the challenge $e_b$
using $x_b, y$ as witness according the protocol $\Sigma_b$. $P$
sends $e_b$, $e_{1-b}$, $z_b)$ and $e_{1-b}$, $z_{1-b}$ to $V$. $V$
checks $e=e_b\oplus e_{1-b}$, and the two transcripts $(a_b, e_b,
z_b)$ and $(a_{1-b}, e_{1-b}, z_{1-b})$ are accepting. The resulting
protocol turns out to be witness indistinguishable: the verifier can
not tell which witness the prover used from a transcript of a
session.

In our rZK argument, the verifier uses a 3-round Witness
Indistinguishable Proof of Knowledge to prove knowledge of one of
the two secret keys associating with his public key. As required in
\cite{DV}, we need a \emph{partial-witness-independence} property
from above proof of knowledge: the message sent at its first round
should have distribution independent from any witness for the
statement to be proved. We can obtain such a protocol using \cite{S}
\cite{DDPY}.

\bigskip\noindent\textbf{Commitment scheme.} A commitment scheme is a
two-phase (committing phase and opening phase) two-party (a sender
$S$ and a receiver $R$)protocol which has following properties: 1)
hiding: two commitments (here we view a commitment as a variable
indexed by the value that the sender committed to) are
computationally distinguishable for every probabilistic
polynomial-time (possibly malicious) $R^*$; 2) Binding: after sent
the commitment to a value $m$, any probabilistic polynomial-time
(possibly malicious) sender $S^*$ cannot open this commitment to
another value $m'\neq m$ except with negligible probability. Under
the assumption of existence of any one-way function families (using
the scheme from \cite{N} and the result from \cite{HILL}) or under
number-theoretic assumptions (e.g., the scheme from \cite{P}), we
can construct a schemes in which the first phase consists of 2
messages. Assuming the existence of one-way permutation families, a
well-known non-interactive (in committing phase) construction of a
commitment scheme (see, e.g. \cite{G}) can be given.

\emph{ A statistically-binding commitment scheme (with computational
hiding)} is a commitment scheme except with a stronger requirement
on binding property: for all powerful sender $S^*$ (without running
time restriction), it cannot open a valid commitment to two
different values except with exponentially small probability. We
refer readers to \cite{G,N} for the details for constructing
statistically-binding commitments.

\emph{A perfect-hiding commitment scheme (with computational
binding)} is the one except with a stronger requirement on hiding
property: the distribution of the commitments is indistinguishable
for all powerful receiver $R^*$. As far as we know, all
perfect-hiding commitment scheme requires interaction (see also
\cite{P,NOVM})in the committing phase.

\begin{definition} \cite{G}. Let $d,r: N\rightarrow N$. we say
that
\[\{f_s: \{0,1\}^{d(|s|)}\rightarrow \{0,1\}^{r(|s|)}\}_{s \in \{0,1\}^*}\]
is an pseudorandom function ensemble if the following two
conditions hold:
\begin{enumerate}
    \item 1. Efficient evaluation: There exists a polynomial-time
    algorithm that on input $s$ and $x\in{{0,1}^{d(|s|)}}$ returns
    $f_s(x)$;
    \item 2. Pseudorandomness: for every probabilistic
    polynomial-time oracle machine $M$, every polynomial
    $p(\cdot)$, and all sufficient large $n's$,
    \[|[Pr[M^{F_n}(1^n)=1]-Pr[M^{H_n}(1^n)=1]|<1/p(n)\]
    where $F_n$ is a random variable uniformly distributed over
    the multi-set $\{f_s\}_{s \in \{0,1\}^n}$, and $H_n$ is uniformly
    distributed among all functions mapping $d(n)$-bit-long
    strings to $r(n)$-bit-long strings.
\end{enumerate}
\end{definition}

\section{A Simple Observation on Resettably-sound Zero Knowledge
Arguments}

resettably-sound zero knowledge argument is a zero knowledge
argument with stronger soundness: for all probabilistic
polynomial-time prover $P^*$, even $P^*$ is allowed to reset the
verifier $V$ to previous state (after resetting the verifier $V$
uses the same random tape), the probability that $P^*$ make $V$
accept a false statement $x\notin L$ is negligible.

In \cite{BGGL} Barak et al. transform a constant round public-coin
zero knowledge argument $<P,V>$ for a $\mathcal{NP}$ language $L$
into a constant round resettably-sound zero knowledge argument
$<P,W>$ for $L$ as follows: equip $W$ with a collection of
pseudorandom functions, and then let $W$ emulate $V$ except that it
generate the current round message by applying a pseudorandom
function to the transcript so far.

 We will use a resettably-sound zero knowledge argument as a
 building block in which the verifier proves to the prover that a
 challenge matches the one that he have committed to in previous
 stage. The simulation for such sub-protocols plays a important
 role in our security reduction, but there is a subtlety in the simulation
 itself.
%The proof of concurrent soundness of our protocol goes by
%contradiction. For the security reduction go through, we need to run
%the simulator for the underlying resettably-sound zero knowledge on
%a false statement.
In the scenario considered in this paper, in which the prover (i.e.,
the verifier in the underlying sub-protocol)can interact with many
copies of the verifier and schedule all sessions at its wish, the
simulation seems problematic because we do not know how to simulate
all the concurrent executions of the Barak's protocol described
below
 \footnote{Barak also presented a constant round bounded concurrent ZK arguments, hence we can obtain a
 constant round resettably-sound bounded concurrent ZK argument by applying the same
 transformation technique to the bounded concurrent ZK argument. We
 stress that in this paper we do not require the bounded concurrent zero knowledge property to hold for the resettably-sound ZK
 argument.}(therefore the resettably-sound zero knowledge
 argument). However, fortunately, it is not necessary to simulate all
 the concurrent executions of the underlying resettably-sound zero knowledge
 argument. Indeed, in order to justify concurrent soundness, we just need to simulate \emph{only one execution} among
 all concurrent executions of the resettably-sound zero knowledge
 argument. We call this property \emph{one-many simulatability}.
 We note that Pass and Rosen \cite{PR} made a similar observation (in a different context) that enables the analysis of
 concurrent non-malleability of their commitment scheme.

 Now we recall the Barak's constant round public-coin
zero knowledge argument \cite{B}, and show this protocol satisfies
\emph{one-many simulatability}, and then so does the
resettably-sound zero knowledge argument transformed from it.

Informally, Barak's protocol for a $\mathcal{NP}$ language $L$
consists of two subprotocol: a general protocol and a WI universal
argument. An real execution of the general protocol generates an
instance that is unlikely in some properly defined language, and in
the WI universal argument the prover proves that the statement $x\in
L$ or the instance generated above is in the properly defined
language. Let $n$ be security parameter and $\{\mathcal{H}_n\}_{n\in
\mathbb{N}}$ be a collection of hash functions where a hash function
$h\in \mathcal{H}_n$ maps $\{0,1\}^*$ to $\{0,1\}^n$, and let
$\textsf{C}$ be a statistically binding commitment scheme. We define
a language $\Lambda$ as follows. We say a triplet $(h,c,r)\in
\mathcal{H}_n\times \{o,1\}^n\times \{o,1\}^n$ is in $\Lambda$, if
there exist a program $\Pi$ and a string $s\in \{0,1\}^{poly(n)}$
such that $z=\textsf{C}(h(\Pi),s)$ and $\Pi(z)=r$ within
superpolynomial time (i.e., $n^{\omega(1)}$).\\

\noindent\textbf{The Barak's Protocol} \cite{B}\\
\textbf{Common input:} an instance $x\in L$ ($|x|=n$)\\
\textbf{Prover's private input:} the witness $w$ such that $(x,w)\in
R_L$\\
$V\rightarrow P$: Send $h\leftarrow_{\small{R}}\mathcal{H}_n$;\\
$P\rightarrow V$: Pick $s\leftarrow_{\small{R}}\{0,1\}^{poly(n)}$
and Send $c=\textsf{C}(h(0^{3n},s)$;\\
$V\rightarrow P$: Send $r\leftarrow_{\small{R}}\{0,1\}^n$;\\
$P\Leftrightarrow V$: A WI universal argument in which $P$ proves
$x\in L$ or $(h,c,r)\in \Lambda$.

\medskip\noindent\textbf{Fact 1.} The
Barak's protocol enjoys \emph{one-many simulatability}. That is, For
every malicious probabilistic polynomial time algorithm $V^*$ that
interacts with (arbitrary) polynomial $s$ copies of $P$ on true
statements $\{x_i\}, 1\leq i\leq s$, and for every $j\in
\{1,2,...,s\}$, there exists a probabilistic polynomial time
algorithm $\textsf{S}$, takes $V^*$ and all witness but the one for
$x_j$, such that the output of $\textsf{S}(V^*,\{(x_i,w_i)\}_{1\leq
i\leq s, i\neq j},x_j)$ (where $(x_i,w_i)\in R_L$) and the view of
$V^*$ are indistinguishable.

\medskip We can construct a simulator
$\textsf{S}=(\textsf{S}_{real},\textsf{S}_j)$ as follows:
$\textsf{S}_{real}$, taking as inputs $\{(x_i,w_i)\}_{1\leq i\leq s,
i\neq j}$, does exactly what the honest provers do on these
statements and outputs the transcript of all but the $j$th sessions
(in $j$th session $x_j\in L$ is to be proven), and $\textsf{S}_j$
acts the same as the simulator associated with Barak's protocol in
the session in which $x_j\in L$ is to be proven, except that when
$\textsf{S}_j$ is required to send a commitment value (the second
round message in Barak's protocol), it commit to the hash value of
the \textbf{joint} residual code of $V^*$ and $\textsf{S}_{real}$ at
this point instead of committing to the hash value of the residual
code of $V^*$ (that is, we treat $\textsf{S}_{real}$ as a subroutine
of $V^*$, and it interacts with $V^*$ internally). We note that the
next message of the joint residual code of $V^*$ and
$\textsf{S}_{real}$ is only determined by the commitment message
from $\textsf{S}_j$, so as showed in \cite{B}, $\textsf{S}_j$ works.
On the other hand, the $\textsf{S}_{real}$'s behavior is identical
to the honest provers. Thus, the whole simulator $\textsf{S}$
satisfies our requirement.

%It seems a little strange to require the simulator get those
%witnesses in a course of simulation. However, this is the case in
%our analysis of concurrent soundness: all the statement (to be
%proven in the underlying resettably sound zero knowledge argument)
%are chosen by the simulator, hence it has all witnesses.

When we transform a constant round public-coin zero knowledge
argument into a resettably-sound zero knowledge argument, the
transformation itself does not influence the simulatability (zero
knowledge) of the latter argument because the zero knowledge
requirement does not refer to the honest verifier (as pointed out in
\cite{BGGL}). Thus, the same simulator described above also works
for the resettably-sound zero knowledge argument in concurrent
settings. So we have

\medskip\noindent\textbf{Fact 2.} The resettably-sound zero knowledge
arguments in \cite{BGGL} enjoy \emph{one-many simulatability}.

\section{rZK Argument with Concurrent Soundness for $\mathcal{NP}$ in the BPK model Under Standard Assumption}
In this section we present a constant-round rZK argument with
concurrent soundness in the BPK model for all $\mathcal{NP}$
language without assuming any subexponential hardness.

For the sake of readability, we give some intuition before
describe the protocol formally.

We construct the argument in the following way: build a concurrent
zero knowledge argument with concurrent soundness and then transform
this argument to a resettable zero knowledge argument with
concurrent soundness. Concurrent zero knowledge with concurrent
soundness was presented in \cite{DV} under standard assumption
(without using "complexity leveraging"). For the sake of
simplification, we modify the \emph{flawed} construction presented
in \cite{Z2} to get concurrent zero knowledge argument with
concurrent soundness. Considering the following two-phase argument
in BPK model: Let $n$ be the security parameter, and $f$ be a one
way function that maps $\{0,1\}^{\kappa(n)}$ to $\{0,1\}^n$ for some
function $\kappa: \mathbb{N}\rightarrow\mathbb{N}$. The verifier
chooses two random numbers $x_0,x_1\in \{0,1\}^{\kappa(n)}$,
computes $y_0=f(x_0)$, $y_1=f(x_1)$ then publishes $y_0$, $y_1$ as
he public key and keep $x_0$ or $x_1$ secret. In phase one of the
argument, the verifier proves to the prover that he knows one of
$x_0$, $x_1$ using a \emph{partial-witness-independently} Witness
Indistinguishable Proof of Knowledge protocol $\Pi_v$. In phase two,
the prover proves that the statement to be proven is true or he
knows one of preimages of $y_0$ and $y_1$ via a witness
indistinguishable argument of knowledge protocol $\Pi_p$. Note that
In phase two we use \emph{argument} of knowledge, this means we
restrict the prover to be a probabilistic polynomial-time algorithm,
and therefore our whole protocol is an argument (not a proof).

Though the above two-phase argument does not enjoy concurrent
soundness \cite{DV}, it is still a good start point and We can use
the same technique in \cite{DV} in spirit to fix the flaw: in phase
two, the prover uses a commitment scheme\footnote{In contrast to
\cite{DV}, we proved that computational binding commitment scheme
suffices to achieve concurrent soundness. In fact, the statistically
binding commitment scheme in \cite{DV} could also be replaced with
computational binding one without violating the concurrent
soundness.}$\textsf{COM}_1$ to compute a commitments to a random
strings $s$, $c=\textsf{COM}_1(s,r)$ ($r$ is a random string needed
in the commitment scheme), and then the prover prove that the
statement to be proven is true or he committed to a preimage of
$y_0$ or $y_1$. We can prove that the modified argument is
concurrent zero knowledge argument with concurrent soundness using
technique similar to that in \cite{DV}.

Given the above (modified) concurrent zero knowledge argument with
concurrent soundness, we can transform it to resettable zero
knowledge argument with concurrent soundness in this way: 1) using a
statistically-binding commitment scheme $\textsf{COM}_0$, the
verifier computes a commitment $c_e=\textsf{COM}_0(e,r_e)$ ($r_e$ is
a random string needed in the scheme) to a random string $e$ in the
phase one, and then he sends $e$ (note that the verifier does not
send $r_e$, namely, it does not open the commitment $c_e$) as the
second message (i.e the challenge) of $\Pi_p$ and prove that $e$ is
the string he committed to in the first phase using resettably sound
zero knowledge argument; 2)equipping the prover with a pseudorandom
function, whenever the random bits is needed in a execution, the
prover applied the pseudorandom function to what he have seen so far
to generate random bits.

Let's Consider concurrent soundness of the above protocol. Imagine
that a malicious prover convince a honest verifier of a false
statement on a session (we call it a cheating session) in an
execution of concurrent attack with high probability. Then we can
use this session to break some hardness assumption: after the first
run of this session, we rewind it to the point where the verifier is
required to send a challenge and chooses an arbitrary challenge and
run the simulator for this underlying resettably-sound zero
knowledge proof. At the end of the second run of this session, we
will extract one of preimages of $y_0$ and $y_1$ from the two
different transcripts, and this contradicts either the witness
indistinguishability of $\Pi_v$ or the binding property of the
commitment scheme $\textsf{COM}_1$. Note that in the above reduction
we just need to simulate the single execution of the
resettably-sound zero knowledge argument in that cheating session,
and do not care about other sessions that initiated by the malicious
prover (in other sessions we play the role of honest verifier). We
have showed the simulation in this special concurrent setting can be
done in a simple way in last section.

%To show the resulting argument enjoys concurrent soundness, we need
%to send a \emph{false} challenge in phase 2 and simulate the
%malicious prover's view. The difficulty of the simulation lies in
%that the malicious prover interacts with the verifier in a
%interleaving way, and we do not know how to construct a constant
%round resettably-sound concurrent zero knowledge argument so far.
%The key observation justifying our analysis is that it is not
%necessary to simulate all the concurrent execution of the underlying
%resettably-sound zero knowledge
% argument, instead, we just need to simulate the \emph{sigle} execution
 %of the underlying resettably-sound zero knowledge argument
% all concurrent executions.\\\\

\medskip\noindent{\textbf{The Protocol (rZK argument with concurrent
soundness in BPK model)}}

\medskip Let $\{prf_r: \{0,1\}^*\rightarrow \{0,1\}^{d(n)}\}_{r\in\{0,1\}^n}$
be a pseudorandom function ensembles, where $d$ is a polynomial
function, $\textsf{COM}_0$ be a \emph{statistically-binding}
commitment scheme, and let $\textsf{COM}_1$ be a general commitment
scheme (can be either statistically-binding or
computational-binding\footnote{If the computational-binding scheme
satisfies perfect-hiding, then this scheme requires stronger
assumption, see also \cite{P,NOVM}}). Without loss of generality, we
assume both the preimage size of the one-way function $f$ and the
message size of $\textsf{COM}_1$ equal $n$.

\textbf{Common input:} the public file $F$, $n$-bit string $x\in L$,
an index $i$ that specifies the $i$-th entry $pk_i=(f,y_0,y_1)$ ($f$
is a one-way function) of $F$.

\textbf{$P$'s Private input:} a witness $w$ for $x\in L$, and a
fixed random string $(r_1,r_2)\in {\{0,1\}^{2n}}$.

\textbf{$V$'s Private input:} a secret key $\alpha$ ($y_0=f(\alpha)$
or $y_1=f(\alpha)$).

\medskip\noindent\textbf{Phase 1:}$V$ Proves Knowledge of $\alpha$ and Sends a Committed Challenge to $P$.

\begin{enumerate}
    \item $V$ and $P$ runs the 3-round
   \emph{partial-witness-independently} witness indistinguishable
   protocol ($\Sigma_{OR}$-protocol) $\Pi_v$ in which $V$ prove knowledge of $\alpha$ that is
   one of the two preimages of $y_0$ and $y_1$. the randomness bits used
   by $P$ equals $r_1$;
    \item $V$ computes $c_e=\textsf{COM}_0(e,r_e)$ for a random $e$ ($r_e$ is a random string
needed in the scheme), and sends $c_e$ to $P$.

\end{enumerate}
\noindent\textbf{Phase 2:} $P$ Proves $x\in L$.

\begin{enumerate}
    \item $P$ checks the transcript of $\Pi_v$ is accepting. if so,
    go to the following step.
    \item $P$ chooses a random string $s, |s|=n$, and compute
    $c=\textsf{COM}_1(s, r_s)$ by picking a randomness $r_s$; $P$ forms a new relation $R'$=$\{(x,y_0,y_1,c,w')\mid (x,w')\in
    R_L \vee (w'=(w{''},r_{w{''}})\wedge y_0=f(w{''})\wedge c=\textsf{COM}_1(w{''},r_{w{''}})) \vee (w'=(w{''},r_{w{''}})\wedge y_1=f(w{''})\wedge
    c=\textsf{COM}_1(w{''},r_{w{''}})))\}$; $P$ invokes the 3-round witness indistinguishable argument of knowledge
($\Sigma_{OR}$-protocol) $\Pi_p$ in which $P$
    prove knowledge of $w'$ such that $(x,y_0,y_1,c;w')\in R'$,
    computes and sends the first message $a$ of $\Pi_p$.\\
    All randomness bits used in this step is obtained by applying the
pseudorandom function
    $prf_{r_2}$ to what $P$ have seen so far, including the common
    inputs, the private inputs and all messages sent by both parties
    so far.
    \item $V$ sends $e$ to $P$, and execute a resettably sound zero
    knowledge argument with $P$ in which $V$ proves to $P$ that
    $\exists$ $r_e$ s.t. $c_e=\textsf{COM}_0(e,r_e)$. Note that the
    subprotocol will costs several (constant) rounds. Again, the
    randomness used by $P$ is generated by applying the pseudorandom function
    $prf_{r_2}$ to what $P$ have seen so far.
    \item $P$ checks the transcript of resettably sound zero knowledge argument
    is accepting. if so, $P$ computes the last message $z$ of
$\Pi_p$ and sends it to $V$.
    \item $V$ accepts if only if $(a,e,z)$ is accepting transcript of $\Pi_p$.
\end{enumerate}

\textbf{Theorem 1.} Let $L$ be a language in $\mathcal{NP}$, If
there exists hash functions collision-resistant against any
polynomial time adversary, then there exists a constant round rZK
argument with concurrent soundness for $L$ in BPK model.

\medskip\textbf{Remark on complexity assumption.} We prove this
theorem by showing the protocol described above is a rZK argument
with concurrent soundness. Indeed, our protocol requires
collision-resistant hash functions and one-way \emph{permutations},
this is because the 3-round $\Sigma$-protocol (therefore
$\Sigma_{OR}$-protocol) for $\mathcal{NP}$ assumes one-way
permutations and the resettably sound zero knowledge argument
assumes collision-resistant hash functions. However, we can build
4-round $\Sigma$-protocol (therefore $\Sigma_{OR}$-protocol) for
$\mathcal{NP}$ assuming existence of one-way functions by adding one
message (see also discussions on $\Sigma$-protocol in section 2),
and our security analysis can be also applied to this variant. We
also note that collision-resistant hash functions implies one-way
functions which suffices to build statistically-binding commitment
scheme \cite{N}(therefore computational-binding scheme), thus, if we
proved our protocol is a rZK argument with concurrent soundness,
then we get theorem 1. Here we adopt the 3-round
$\Sigma_{OR}$-protocol just for the sake of simplicity.

\bigskip\emph{Proof.} \textbf{Completeness.} Straightforward.

\textbf{Resettable (\emph{black-box}) Zero Knowledge.} The analysis
is very similar to the analysis presented in \cite{CGGM,DPV2}. Here
we omit the tedious proof and just provide some intuition. As usual,
we can construct a simulator $\textsf{Sim}$ that extracts all secret
keys corresponding to those public keys registered by the malicious
verifier from $\Pi_v$ and then uses them as witness in executions of
$\Pi_p$, and $\textsf{Sim}$ can complete the simulation in expected
polynomial time. We first note that when a malicious verifier resets
a an honest prover, it can not send two different challenge for a
fixed commitment sent in Phase 1 to the latter because of
statistically-binding property of $\textsf{COM}_0$ and resettable
soundness of the underlying sub-protocol used by the verifier to
prove the challenge matches the value it has committed to in Phase
1. To prove the property of rZK, we need to show that the output of
$\textsf{Sim}$ is indistinguishable form the real interactions. This
can be done by constructing a non-uniform hybrid simulator
$\textsf{HSim}$ and showing the output of $\textsf{HSim}$ is
indistinguishable from both the output of $\textsf{Sim}$ and the
real interaction. $\textsf{HSim}$ runs as follows. Taking as inputs
all these secret keys and all the witnesses of statements in
interactions, $\textsf{HSim}$ computes commitments exactly as
$\textsf{Sim}$ does but executes $\Pi_p$ using the same witness of
the statement used by the honest prover. It is easy to see that the
output of the hybrid simulator is indistinguishable from both the
transcripts of real interactions (because of the
computational-hiding property of $\textsf{COM}_1$) and the output of
$\textsf{Sim}$ (because of the witness indistinguishability of
$\Pi_p$), therefore, we proved the the output of $\textsf{Sim}$ is
indistinguishable form the real interactions.

\textbf{Concurrent Soundness.} Proof proceeds by contradiction.
%The
%techniques used here is similar to but different from that in
%\cite{DV}.

Assume that the protocol does not satisfy the concurrent soundness
property, thus there is a $s$-concurrently malicious prover $P^*$,
concurrently interacting with $V$, makes the verifier accept a false
statement $x\notin L$ in $j$th session with non-negligible
probability $p$.

We now construct an algorithm $\textsf{B}$ that takes the code (with
randomness hardwired in)of $P^*$ as input and breaks the one-wayness
of $f$ with non-negligible probability.

$\textsf{B}$ runs as follows. On input the challenge $f,y$ (i.e.,
given description of one-way function, $\textsf{B}$ finds the
preimage of $y$), $\textsf{B}$ randomly chooses $\alpha\in
\{0,1\}^n$, $b\in {\{0,1\}}$, and guess a session number
$j\in{\{1,...,s\}}$(guess a session in which $P^*$ will cheat the
verifier successfully on a false statement $x$. Note that the event
that this guess is correct happens with probability $1/s$), then $B$
registers $pk=(f,y_0,y_1)$ as the public key, where $y_b=f(\alpha)$,
$y_{1-b}=y$. For convenience we let $x_b=\alpha$, and denote by
$x_{1-b}$ one of preimages of $y_{1-b}$ ($y_{1-b}=y=f(x_{1-b})$).
Our goal is to find one preimage of $y_{1-b}$.

We write $\textsf{B}$ as
$\textsf{B}=(\textsf{B}_{real},\textsf{B}_j)$. $\textsf{B}$
interacts with $P^*$ as honest verifier (note that $\textsf{B}$
knows the secret key $\alpha$ corresponding the public key $pk$) for
all but $j$th session. Specifically, $\textsf{B}$ employs the
following extraction strategy:
\begin{enumerate}
    \item $\textsf{B}$ acts as the honest verifier in this stage.
     That is, it completes $\Pi_v$ using $\alpha=x_b$ as
secret key, and commits to $e$, $c_e=\textsf{COM}_0(e,r_e)$ in phase
1 then runs resettably sound ZK argument in Phase 2 using $e$, $r_e$
as the witness. In particular, $\textsf{B}$ uses $\textsf{B}_j$ to
play the role of verifier in the $j$th session, and uses
$\textsf{B}_{real}$ to play the role of verifier in all other
sessions. At the end of $j$th session, if $B$ gets an accepting
transcript $(a,e,z)$ of $\Pi_p$, it enters the following rewinding
stage; otherwise, $B$ halts and output $"\bot"$
    \item $\textsf{B}_j$ rewind $P^*$ to the point of beginning of step
    3 in Phase 2 in $j$th session, it chooses a random string $e'\neq e$
and simulates the underlying resettably sound ZK argument in the
same way showed in section 3: it commits to the hash value of the
joint residual code of $P^*$ and $\textsf{B}_{real}$ in the second
round of the resettably sound ZK argument (note this subprotocol is
transformed from Barak's protocol) and uses them as the witness to
complete the proof for the following \emph{false} statement:
$\exists$ $r_e$ s.t. $c_e=\textsf{COM}_0(e',r_e)$. If this rewinds
incurs some other rewinds on other sessions, $\textsf{B}_{real}$
always acts as an honest verifier. When $\textsf{B}$ get another
accepting transcript $(a,e',z')$ of $\Pi_p$ at step 5 in Phase 2 in
$j$th session, it halts, computes the witness from the two
transcripts and outputs it, otherwise, $\textsf{B}$ plays step 3 in
$j$th session again.
\end{enumerate}

We denote this extraction with $\emph{Extra}$.

We first note that $\textsf{B}$'s simulation of $P^*$'s view only
differs from $P^*$'s view in real interaction with an honest
verifier in the following: In the second run of $\Pi_p$ in $j$th
session $\textsf{B}$ proves a \emph{false} statement to $P^*$ via
the resettably sound zero knowledge argument instead of executing
this sub-protocol honestly. We will show that this difference is
computationally indistinguishable by $P^*$ using the technique
presented in the analysis of resettable zero knowledge property, or
otherwise we can use $P^*$ to violate the zero knowledge property of
the underlying resettably sound zero knowledge argument or the
statistically-binding property of the commitment scheme
$\textsf{COM}_0$. We also note that if the simulation is successful,
$\textsf{B}$ gets an accepting transcript of $\Pi_p$ in stage 1 with
probability negligibly close to $p$, and once $\textsf{B}$ enters
the rewinding stage (stage 2) it will obtain another accepting
transcript in expected polynomial time because $p$ is
non-negligible. In another words, $\textsf{B}$ can outputs a valid
witness with probability negligibly close to $p$ in the above
extraction.

Now assume $\textsf{B}$ outputs a valid witness $w'$ such that
$(x,y_0,y_1,c,w')\in R'$, furthermore, the witness $w'$ must satisfy
$w'=(w{''},r_{w{''}})$ and $y_b=f(w{''})$ or $y_{1-b}=f(w{''})$
because $x\notin L$. If $y_{1-b}=f(w{''})$, we break the one-way
assumption of $f$ (find the one preimage of $y_{1-b}$),
otherwise(i.e., $w{''}$ satisfies $y_b=f(w{''})$), we fails. Next we
claim $\textsf{B}$ succeed in breaking the one-way assumption of $f$
with non-negligible probability.

Assume otherwise, with at most a negligible probability $q$,
$\textsf{B}$ outputs one preimage of $y_{1-b}$. Then We can
construct a non-uniform algorithm $\textsf{B'}$ (incorporating the
code of $P^*$)to break the witness indistinguishability of $\Pi_v$
or the computational binding of the commitment scheme
$\textsf{COM}_1$.

The non-uniform algorithm $\textsf{B'}$ takes as auxiliary input
$(y_0, y_1, x_0, x_1)$ (with input both secret keys) and interacts
with $P^*$ under the public key $(y_0, y_1)$. It performs the
following experiment:

\begin{enumerate}
    \item \emph{Simulation} (\emph{until} $\textsf{B'}$ \emph{receives
    the first message $a$ of $\Pi_p$ in $j$th session}). $\textsf{B'}$
    acts exactly as the $\textsf{B}$. Without
    loss of generality, let $\textsf{B'}$ uses $x_0$ as
    witness in all executions of $\Pi_v$ that completed before step 2 in Phase 2 of the $j$th session.
    Once $\textsf{B'}$ receives
    the first message $a$ of $\Pi_p$ in $j$th session, it splits
    this
    experiment and continues independently in following games:
    \item \emph{Extracting Game 0}. $\textsf{B'}$ continues the above simulation
    and uses the same extraction strategy of $\textsf{B}$. In particular,
    it runs as follows. 1) continuing to simulate: $\textsf{B}$ uses $x_0$ as witness in all
executions of $\Pi_v$ that take place during this game; 2)
extracting: if $\textsf{B}$ obtained an accepting transcript $(a,
e_0, z_0)$ at the end of the first run of $\Pi_p$ in $j$th session,
it rewinds to the point of beginning of
    step 3 in Phase 2 in $j$th session and replays this round
by sending another random challenge $e'\neq e$ until he gets another
accepting transcript $(a, e_0', z_0')$ of $\Pi_p$, and then
$\textsf{B}$ outputs a valid witness, otherwise outputs $"\bot"$.
    \item \emph{Extracting Game 1}: $\textsf{B'}$ repeats Extracting Game 0 but $\textsf{B'}$ uses $x_1$ as
witness in all executions of $\Pi_v$ during this game (i.e., those
executions of $\Pi_v$ completed after the step 2 in Phase 2 in the
$j$th session). At the end of this game, $\textsf{B'}$ either
obtains two accepting transcripts $(a, e_1, z_1)$, $(a, e_1', z_1')$
and outputs an valid witness, or outputs $"\bot"$. Note that an
execution of $\Pi_v$ that takes place during this game means at
least the last (third) message of $\Pi_v$ in that execution has not
yet been sent before step 2 in Phase 2 in $j$th session. Since the
$\Pi_v$ is \emph{partial-witness-independent} $\Sigma$-protocol (so
we can decide to use which witness at the last (third) step of
$\Pi_v$), $\textsf{B'}$ can choose witness at its desire to complete
that execution of $\Pi_v$ after the step 2 in Phase 2 in the $j$th
session.
\end{enumerate}

We denote by $\emph{EXP}_0$ the \emph{Simulation} in stage 1
described above with its first continuation \emph{Extracting Game
0}, similarly, denote by $\emph{EXP}_1$ the same \emph{Simulation}
with its second continuation \emph{Extracting Game 1}.

Note that the $P^*$'s view in $\emph{EXP}_0$ is identical to its
view in $\emph{EXTRA}$ in which $\textsf{B}$ uses $x_0$ ($b=0$)as
witness in all executions of $\Pi_v$, so the outputs of
$\textsf{B'}$ at the end of $\emph{EXP}_0$ is identical to the
outputs of $\textsf{B}$ taking $x_0$ as the secret key in
$\emph{EXTRA}$, that is, with non-negligible probability $p$
$\textsf{B'}$ outputs one preimage of $y_0$, and with negligible
probability $q$ it outputs one preimage of $y_1$.

Consider $\textsf{B}$'s behavior in $\emph{EXTRA}$ when it uses
$x_1$($b=1$)as the secret key. The behavior of $\textsf{B}$ only
differs from the behavior of $\textsf{B'}$ in $\emph{EXP}_1$ in
those executions of $\Pi_v$ that completed before the step 2 in
Phase 2 in the $j$th session: $\textsf{B'}$ uses $x_0$ as witness in
all those executions, while $\textsf{B}$ uses $x_1$ as witness.
However, the $P^*$ cannot tell these apart because $\Pi_v$ is
witness indistinguishable and all those executions of $\Pi_v$ have
not been rewound during both $\emph{EXTRA}$ and $\emph{EXP}_1$ (note
that $\textsf{B'}$ does not rewind past the the step 2 in Phase 2 in
the $j$th session in the whole experiment). Thus, we can claim that
at the end of $\emph{EXP}_1$, $\textsf{B'}$ outputs one preimage of
$y_1$ with probability negligibly close to $p$, and it outputs one
preimage of $y_0$ with probability negligibly close to $q$.

In the above experiment conducted by $\textsf{B}$, the first message
$a$ sent by $P^*$ in the $j$th session contains a commitment $c$ and
this message $a$ (therefore $c$) remains unchanged during the above
whole experiment. Clearly, with probability negligibly close to
$p^2$ (note that $q$ is negligible), $\textsf{B'}$ will output two
valid witness $w_0'=(w_0{''},r_{w_0{''}})$ and
$w_1'=(w_1{''},r_{w_1{''}})$ (note that $w_0{''}\neq w_1{''}$ except
for a very small probability) from the above two games such that the
following holds: $y_0=f(w_0{''})$, $y_1=f(w_1{''})$,
$c=\textsf{COM}_1(w_0{''},r_{w_0{''}})$ and
$c=\textsf{COM}_1(w_1{''},r_{w_1{''}})$. This contradicts the
computational-binding property of the scheme $\textsf{COM}_1$.

In sum, we proved that if $\textsf{COM}_1$ enjoys
computational-binding and $\Pi_v$ is witness indistinguishable
protocol with \emph{partial-witness-independence} property, then
$\textsf{B}$ succeeds in breaking the one-wayness of $f$ with
non-negligible probability. In another words, if the one-way
assumption on $f$ holds, it is infeasible for $P^*$ to cheat
an honest verifier in concurrent settings with non-negligible probability. $\Box$\\

\bigskip\noindent\textbf{Acknowledgments.} Yi Deng thanks
Giovanni Di Crescenzo, Rafael Pass, Ivan Visconti and Yunlei Zhao
for many helpful discussions and classifications.

\end{document}